\documentstyle[preprint,aps]{revtex}
\begin{document}
\draft
\preprint{\hbox to \hsize{\hfil\vtop{\hbox{IASSNS-HEP-96/37}
\hbox{HUB-EP-96/9}
\hbox{April, 1996}}}}

\title{Photon Splitting in a Strong Magnetic Field:
Recalculation and Comparison With Previous Calculations\\
}
\author{Stephen L. Adler\\}     
\address{
Institute for Advanced Study\\
Princeton, NJ 08540\\
}
\author{Christian Schubert\\}     
\address{
Humboldt Universitit\"at zu Berlin\\
Invalidenstr.  110, D-10115 Berlin, Germany\\
}
\date{\today}
\maketitle
\begin{abstract}
We recalculate the amplitude for photon splitting in a strong magnetic
field below the pair production threshold, using the worldline path 
integral variant of the Bern--Kosower formalism.  Numerical comparison 
(using programs that we have made available for public access 
on the Internet) shows that the results of the recalculation are identical 
to the earlier calculations of Adler and later of Stoneham, and to the recent 
recalculation by Baier, Milstein, and Shaisultanov.  
\end{abstract}
\pacs{
{\tt$\backslash$\string pacs\{12.20.Ds, 95.30.Cq\}}}
\narrowtext
\section*{}
Photon splitting in a strong magnetic field is an interesting process, both 
from a theoretical viewpoint because of the relatively sophisticated methods 
needed to do the calculation, and because of its potential astrophysical 
applications.   The first calculation  to exactly include the corrections 
arising from nonzero photon frequency $\omega$ was given by Adler [1], 
who obtained the amplitude as a triple integral that is strongly 
convergent below the pair production threshold at $\omega=2m$, 
and who included a numerical evaluation for the special case $\omega=m$.
Subsequently, the calculation was repeated by Stoneham [2] using a different 
method, leading to 
a different expression as a triple integral, that has never been compared 
to the formula of Ref. [1] either analytically or numerically.  Recently, 
a new calculation has been published by Mentzel, Berg, and Wunner [3] in the 
form of a triple infinite sum, and numerical evaluation of their formula by 
Wunner, Sang, and Berg [4] claims photon splitting rates roughly four orders 
of magnitude larger than those found in Ref. [1].  Since this result, if 
correct, would have important astrophysical implications, a recalculation 
by an independent method seems in order.  We report the results of such 
a recalculation here, together with a numerical comparison of the resulting 
amplitude with those of Adler and of Stoneham, as well as with a recent 
recalculation independently carried out by Baier, Milstein, and 
Shaisultanov [5].  The comparison shows that these four 
independent calculations 
give precisely the same amplitude, showing no evidence of the dramatic 
energy dependent effects claimed in Refs. [3] and [4]. 

Our recalculation of the photon splitting amplitude 
uses a variant of the 
worldline path integral approach to the 
Bern--Kosower formalism
[6,~7,~8,~9].
As is well known, the
one loop QED effective action induced for the photon field
by a spinor loop
can be represented by the following double path integral, 

\begin{eqnarray}
\Gamma\lbrack A\rbrack &  = &- 2 {\displaystyle\int_0^{\infty}}
{ds\over s}
e^{-m^2s}
{\displaystyle\int} {\cal D} x\cal D\psi\nonumber\\
& \phantom{=}
&\times {\rm exp}\biggl [- \int_0^s d\tau
\Bigl ({1\over 4}{\dot x}^2 + {1\over
2}\psi\dot\psi
+ ieA_{\mu}\dot x^{\mu} - ie
\psi^{\mu}F_{\mu\nu}\psi^{\nu}
\Bigr )\biggr ].
\label{one}
\end{eqnarray}
\noindent
Here $s$ is the usual Schwinger proper--time parameter,
the $x^{\mu}(\tau )$'s are the
periodic functions from the circle
with circumference $s$ into 
spacetime, and
the $\psi^{\mu}(\tau )$'s are
antiperiodic and Grassmann--valued.

Photon scattering amplitudes are obtained by
specializing the 
background to a sum of plane waves with
definite polarizations. Both path integrals are then
evaluated by one--dimensional perturbation theory,
i.e. one obtains an integral representation for
the $N$--photon amplitude
by Wick--contracting $N$ ``photon vertex operators''

\begin{equation}
V=
\int_0^T d\tau\,\bigl[\dot x^{\mu}\varepsilon_{\mu}
-2{\rm i}\psi^{\mu}\psi^{\nu}k_{\mu}\varepsilon_{\nu}\bigr]
{\rm exp}[ikx({\tau})].
\label{two}
\end{equation} 
\noindent
The appropriate one--dimensional propagators are
\begin{eqnarray}
\langle y^{\mu}(\tau_1)y^{\nu}(\tau_2)\rangle
   & = &- g^{\mu\nu}G_B(\tau_1,\tau_2)
    = \quad - g^{\mu\nu}\biggl[ \mid \tau_1 - \tau_2\mid -
{{(\tau_1 - \tau_2)}^2\over s}\biggr],\nonumber\\
\langle \psi^{\mu}(\tau_1)\psi^{\nu}(\tau_2)\rangle
   & = &{1\over 2}\, g^{\mu\nu} G_F(\tau_1,\tau_2)
   = \quad {1\over 2}\, g^{\mu\nu}{\rm sign}
(\tau_1 -\tau_2 )\, .
\label{three}
\end{eqnarray}
\noindent
The bosonic Wick contraction is actually carried out in the relative coordinate
$y(\tau)=x(\tau)-x_0$ of the closed loop, while
the (ordinary) integration over the average
position $x_0={1\over s}\int_0^sd\tau x(\tau)$
yields energy--momentum conservation.

To take the additional constant magnetic background field 
$B$ into account, one chooses Fock--Schwinger gauge, where its 
contribution to the worldline Lagrangian becomes

\begin{equation}
\Delta{\cal L} = {1\over 2}iey^{\mu} F_{\mu\nu}
\dot y^{\nu} - ie\psi^{\mu} F_{\mu\nu}\psi^{\nu}\quad .
\label{four}
\end{equation}
\noindent
Being bilinear, those terms can be simply
absorbed into the kinetic part of the  
Lagrangian 
[9~,10].  This leads to generalized
worldline propagators defined by

\begin{eqnarray}
{1\over 2}\biggl(
{\partial^2\over {\partial\tau}^2}
-2ieF{\partial\over {\partial\tau}}
\biggr){{\cal G}_B}(\tau_1,\tau_2) 
&=& \delta(\tau_1 -\tau_2) 
-{1\over s},
\label{five}\\
{1\over 2}\biggl(
{\partial\over {\partial\tau}}
-2ieF\biggr)
{{\cal G}_F}(\tau_1,\tau_2) &=& 
\delta(\tau_1 -\tau_2). 
\label{six}
\end{eqnarray}
\noindent
The solutions to these equations 
can be written in the form [11]

\begin{eqnarray}
{{\cal G}_B}(\tau_1,\tau_2) &=&
{1\over 2{(eF)}^2}\biggl({eF\over{{\rm sin}(esF)}}
{\rm e}^{-iesF\dot G_{B12}}
+ieF\dot G_{B12} -{1\over s}\biggr),
\label{seven}\\
{\cal G}_{F}(\tau_1,\tau_2) &=&
G_{F12}
{{\rm e}^{-iesF\dot G_{B12}}\over {\rm cos}(esF)}\,,
\label{eight}
\end{eqnarray}

\noindent
(we have abbreviated $G_{Bij}:=G_B(\tau_i,\tau_j)$, and a dot always
denotes a derivative with respect to the first variable).
Those expressions should be understood as power series in the
field strength matrix.
To obtain the photon splitting amplitude, we will 
use them for the Wick contraction of
three vertex operators $V_0$ and $V_{1,2}$, representing the
incoming and the two outgoing photons. 

The calculation is
greatly simplified by the special kinematics of this process.
Energy--momentum conservation $k_0+k_1+k_2=0$
forces collinearity of all three four--momenta, so that,
writing $-k_0\equiv k\equiv \omega n$,

\begin{equation}
k_1={\omega_1\over\omega}k,k_2={\omega_2\over\omega}k;k^2=k_1^2
=k_2^2=k\cdot k_1=k\cdot k_2=k_1\cdot k_2=0.
\label{nine}
\end{equation}
\noindent
Moreover, a simple CP--invariance argument together with an analysis of 
dispersive effects [1] shows that 
there is only one allowed polarization case. This is the one
where the incoming photon is polarized parallel to the plane
containing the external field and the direction of propagation,
and both outgoing ones are polarized perpendicular to this plane. This choice
of polarizations leads to the
further vanishing relations

\begin{equation}
\varepsilon_{1,2}\cdot\varepsilon_0
=\varepsilon_{1,2}\cdot k
=\varepsilon_{1,2}\cdot F=0\quad .
\label{ten}
\end{equation}
\noindent
In particular, we cannot Lorentz contract $\varepsilon_1$ with
anything but $\varepsilon_2$. This leaves us with only a small
number of nonvanishing Wick contractions:

\begin{eqnarray}
&&\langle V_0V_1V_2\rangle = i\exp\biggl[{1\over 2}
\sum_{i,j=0}^2
\bar\omega_i\bar\omega_j
n{\cal G}_{Bij}n\biggr]
\Biggl\lbrace\Bigl[
\varepsilon_1\ddot{\cal G}_{B12}\varepsilon_2
+\varepsilon_1{\cal G}_{F12}\varepsilon_2
\bar\omega_1\bar\omega_2
n{\cal G}_{F12}n\Bigr]\nonumber\\
&&\times\Bigl[-\sum_{i=0}^2
\bar\omega_i\varepsilon_0
\dot{\cal G}_{B0i}n
+\bar\omega_0\varepsilon_0{\cal G}_{F00}n\Bigr]
-\bar\omega_0\bar\omega_1\bar\omega_2
\varepsilon_1{\cal G}_{F12}
\varepsilon_2\Bigl[n{\cal G}_{F10}\varepsilon_0
\,n{\cal G}_{F20}n-(1\leftrightarrow 2)\Bigl]
\Biggr\rbrace. 
\label{eleven}
\end{eqnarray}
\noindent
For compact notation we have defined
$\bar\omega_0=\omega, \bar\omega_{1,2}=-\omega_{1,2}$.
This result has still to be multiplied by an overall factor of
$(esB)\cosh(esB)\over {(4\pi s)}^2{\sinh(esB)}$,
which by itself would just produce the
Euler--Heisenberg Lagrangian, and here
appears as the product of the two free Gaussian
path integrals [8].

It is then a matter of simple algebra to
obtain the following representation for the
matrix element $C_2[\omega,\omega_1,\omega_2,B]$ 
appearing in Eq.~(25) of [1]:

\begin{eqnarray}
&&C_2[\omega,\omega_1,\omega_2,B] =
{m^8\over 4 \omega\omega_1\omega_2}
\int_0^{\infty}dss{{\rm e}^{-m^2s}
\over {(esB)}^2{\rm sinh}(esB)}
\int_0^sd\tau_1\int_0^sd\tau_2\nonumber\\
&&\times{\rm exp}
\biggl\lbrace-{1\over 2}\sum_{i,j=0}^2\bar\omega_i
\bar\omega_j\Bigl[G_{Bij}+{1\over 2eB}
{{\rm cosh}(esB\dot G_{Bij})\over {\rm sinh}(esB)}
\Bigr]\biggr\rbrace\nonumber\\
&&\times 
\Biggl\lbrace\biggl\lbrack
-\cosh (esB)\ddot G_{B12} +\omega_1\omega_2
\Bigl( {\rm cosh}(esB)-\cosh (esB\dot G_{B12})\Bigr )\biggr\rbrack
\nonumber\\
&&\times \biggl\lbrack
\omega\Bigl(\coth (esB)-\tanh (esB)\Bigr)
-\omega_1
{{\rm cosh}(esB\dot G_{B01})\over {\rm sinh}(esB)}
-\omega_2
{{\rm cosh}(esB\dot G_{B02})\over {\rm sinh}(esB)}
\biggr\rbrack\nonumber\\
&& +\omega\omega_1\omega_2
{G_{F12}\over \cosh (esB)}
\biggl\lbrack
\sinh (esB\dot G_{B01})\Bigl (\cosh (esB)-\cosh (esB\dot G_{B02})
\Bigr ) - 
\Bigl ( 1\leftrightarrow 2\Bigr )
\biggr\rbrack\Biggl\rbrace.
\label{twelve}
\end{eqnarray}

\noindent
Here translation invariance in $\tau$ has been used to
set the position $\tau_0$ of the
incoming photon equal to $s$. 
Coincidence limits have to be treated according to the rules
$\dot G_{B}(\tau,\tau)=0, \dot G_{B}^2(\tau,\tau)=1$.

Alternatively, one may
remove $\ddot G_{B12}$ 
by partial integration on the circle. This leads to the
equivalent formula 

\begin{eqnarray}
&&C_2[\omega,\omega_1,\omega_2,B] =
{m^8\over 4}
\int_0^{\infty}dss\,
{\rm e}^{-m^2s}{\cosh (esB)\over{{(esB)}^2 {\rm sinh}(esB)}}
\int_0^sd\tau_1\int_0^sd\tau_2\nonumber\\
&&\times{\rm exp}
\biggl\lbrace-{1\over 2}\sum_{i,j=0}^2\bar\omega_i
\bar\omega_j\Bigl[G_{Bij}+{1\over 2eB}
{{\rm cosh}(esB\dot G_{Bij})\over {\rm sinh}(esB)}
\Bigr]\biggr\rbrace\nonumber\\
&&\times 
\Biggl\lbrace
\biggl\lbrack
\dot G_{B12}\Bigl(\dot G_{B12}-
{\sinh(esB\dot G_{B12})\over \sinh(esB)}\Bigr)
-\Bigl(1-{\cosh(esB\dot G_{B12})\over \cosh(esB)}\Bigr)
\biggr\rbrack\nonumber\\&&
\times
\biggr\lbrack-\coth (esB)+\tanh (esB)
+{\omega_1\over \omega}{\cosh(esB\dot G_{B01})\over \sinh(esB)}
+{\omega_2\over \omega}
{\cosh(esB\dot G_{B02})\over \sinh(esB)}\biggr\rbrack
\nonumber\\&&
+\dot G_{B12}
\biggl\lbrack
\Bigl({{\rm cosh}(esB\dot G_{B02})\over {\rm sinh}(esB)}
-{1\over esB}\Bigr)
\Bigl(\dot G_{B01}
-{{\rm sinh}(esB\dot G_{B01})\over {\rm sinh}(esB)}
\Bigr)
- \Bigl( 1\leftrightarrow 2\Bigr )
\biggl\rbrack
\nonumber\\&&
+{1\over 2}
\dot G_{B12}
\biggl\lbrack{\omega\over \omega_2}
\Bigl(\dot G_{B01}-{\sinh (esB\dot G_{B01})\over \sinh (esB)}
\Bigr)
-\Bigl( 1\leftrightarrow 2\Bigr)
\biggr\rbrack
\Bigl(
-\coth (esB)+{1\over esB}+\tanh (esB)\Bigr)
\nonumber\\
&&+ G_{F12}
\biggl\lbrack
{\sinh (esB\dot G_{B01})
\over \cosh(esB)}
\Bigl( 1-{\cosh (esB\dot G_{B02})\over \cosh(esB)}
\Bigr )
-\Bigl( 1\leftrightarrow 2\Bigr)
\biggr\rbrack
\Biggl\rbrace.
\label{thirteen}
\end{eqnarray}

\noindent
This form of the amplitude
is less compact, but the integrand (apart from the exponential) 
is homogeneous in the
$\omega_i$.

Finally, let us remark that
the analogous expression for scalar QED would be 
obtained by deleting 
all terms in Eq.~(11)
containing a ${\cal G}_F$, as well as
the $\cosh(esB)$ appearing in the overall
factor and
the global factor of $-2$ in Eq.~(1).

In order to compare the amplitudes of Eqs.~(12) and (13) to those of 
Refs.~[1], [2], and [5], we observe that both Eq.~(12) and Eq.~(13) 
can be written in the form
\begin{equation}
C_2[\omega,\omega_1,\omega_2,B]={m^8 \over 4 B^2 \omega \omega_1 \omega_2}
\int_0^{\infty} {ds \over s} e^{-m^2 s} J_2(s,\omega,\omega_1,\omega_2,B)~,
\label{fourteen}
\end{equation}
in which $J_2$ is independent of the electron mass $m$.  Inspection 
shows that the amplitude expressions of Adler [1] and Baier, Milstein, 
and Shaisultanov [5] are already in the form of Eq.~(14), while that 
of Stoneham [2] can be put in this form by doing an integration by 
parts in the proper time parameter $s$, using the identity 
\begin{equation}
m^2 e^{-m^2 s} =-{d \over ds} e^{-m^2s}~~~  
\label{fifteen}
\end{equation}
to eliminate a term proportional to $m^2$ in the amplitude.  
In rewriting Stoneham's formulas in this form, we note that his 
$M_1(B)$ is what we are calling $C_2[\omega,\omega_1,\omega_2,B]$, and that
there is an error of an overall minus sign in either his Eq.~(37) or 
the first line of his Eq.~(40).
Similarly, in rewriting the formulas of Baier, Milstein, and Shaisultanov 
in this form, we note that their amplitude $T$ is related to $C_2$ by 
\begin{equation}
C_2[\omega,\omega_1,\omega_2,B]={\pi^{1\over2} m^8 \over 4 \alpha^3 B^3 
\omega \omega_1 \omega_2} T ~~~. 
\label{sixteen}
\end{equation}

Once all amplitudes are put in the form of Eq.~(14), we can compare them 
by comparing the proper time integrand $J_2(s,\omega,\omega_1,\omega_2,B)$, 
which in each case involves only a double integral over a bounded domain.  
The only remaining subtlety is that we must remember that $J_2$ vanishes 
as $\omega \omega_1 \omega_2$ for small photon energy; this is manifest in 
Eq.~(13) above, but in Eq.~(12) and the corresponding equations obtained 
from Refs.~[1], [2], and [5], there is an apparent linear term in the 
frequencies which vanishes when the double integral is done exactly.  In 
order to get robust results for small photon frequency when the double 
integral is done numerically, this linear term must first be subtracted 
away, by replacing expressions of the form 

\begin{mathletters}
\label{allequations} 
\begin{equation}
\int \int e^Q (L +C)  
\label{equationa}
\end{equation}
with $L$, $Q$, and $C$ respectively linear, quadratic, and cubic in the 
photon frequencies, by  the subtracted expression 
\begin{equation}
\int \int [(e^Q-1)L+C]  ~~~.
\label{equationb}
\end{equation}
\end{mathletters}
This subtraction is already present in the expression of Eq.~(25) of 
Ref. [1], and is discussed in the form of Eqs.~(17a, b) in Ref.~[5], and 
it also must be applied to Eqs.~(37) and (39) of Ref.~[2] after the 
integration by parts of Eq.~(15) has been carried out.  While in principle 
this subtraction should be applied to Eq.~(12) above, it turns out not to 
be needed there, because the linear term in the frequencies involves only 
integrals of the general form
\begin{equation}
\int_0^s d\tau_1 f(s,\tau_1) \int_0^s d\tau_2 
[\delta(\tau_1-\tau_2) -1/s]~~~,
\label{eighteen}
\end{equation}
which is exactly zero using a discrete trapezoidal integration method when 
the $\delta$ function is discretized as a Kronecker delta.  Thus Eq.~(12)
is robust for small photon frequencies as it stands, when used in 
conjunction with trapezoidal integration.

With these preliminaries out of the way, it is then completely 
straightforward to program the functions $J_2(s,\omega,\omega_1,\omega_2,B)$ 
for the five cases represented by the formulas of Adler [2], Stoneham [3], 
Eq.~(12) of this paper, Eq.~(13) of this paper, and Baier, Milstein, and
Shaisultanov [5], with the result that they are all seen to be precisely 
the same; the residual errors approach zero quadratically as the integration 
mesh spacing approaches zero, as expected for trapezoidal integration.  
We have not carried out the $s$ and $\omega_1$ integrals needed to get the 
photon splitting absorption coefficient, since this was done in Ref.~[1],  
with results confirmed by the more extensive numerical analysis given 
in Ref.~[5].  However, anyone wishing to do this further computation 
can obtain our programs for calculating the proper time integrand 
$J_2$ by accessing S. L. A.'s home page on the 
Institute for Advanced Study web site
(http://www.sns.ias.edu/$\sim$adler/Html/photonsplit.html).
 
\acknowledgments
C.S. would like to thank P. Haberl for help with numerical
work, and the DFG for financial support. 
S.L.A. wishes to acknowledge the hospitality 
of the Institute for Theoretical Physics in Santa Barbara, where parts 
of this work were done.  He also wishes to thank J.N. Bahcall for suggesting
that this work be undertaken, and V.N. Baier for informing him of the 
results of Ref. [5]. 
This work was supported in part by the Department of Energy under
Grant \#DE--FG02--90ER40542.

\end{document}